# Genetic Algorithm to Make Persistent Security and Quality of Image in Steganography from RS Analysis

T. R. Gopalakrishnan Nair#[1], Suma V[#2], Manas S[#3]

[1,2]*Research and Industry Incubation Center, Dayananda Sagar Institutions, Bengaluru, India*
[1]<trgnair@gmail.com>, [2]sumavdsce@gmail.com
[3]*Dayananda Sagar College of Engineering, Bengaluru, India*
[3]manas.maanu@gmail.com

*Abstract—* **Retention of secrecy is one of the significant features during communication activity. Steganography is one of the popular methods to achieve secret communication between sender and receiver by hiding message in any form of cover media such as an audio, video, text, images etc. Least significant bit encoding is the simplest encoding method used by many steganography programs to hide secret message in 24bit, 8bit colour images and grayscale images. Steganalysis is a method of detecting secret message hidden in a cover media using steganography. RS steganalysis is one of the most reliable steganalysis which performs statistical analysis of the pixels to successfully detect the hidden message in an image. However, existing steganography method protects the information against RS steganalysis in grey scale images. This paper presents a steganography method using genetic algorithm to protect against the RS attack in colour images. Stego image is divided into number of blocks. Subsequently, with the implementation of natural evolution on the stego image using genetic algorithm enables to achieve optimized security and image quality.**

## 1 INTRODUCTION

Steganography is a method of hiding a secret message in any cover media. Cover media can be a text, or an image, an audio or video etc. Several techniques were followed to maintain secrecy of the communication between sender and receiver. The techniques include sending of the secret message in a tattooed form on the shaved slave head. The slave was sent to the recipient only after the growth of his hair. Another method of hiding secret message was to write on a wooden board and to cover with wax [14]. Steganography is a method of hiding a message where only the sender and recipient are aware of existence of the message. The Greek word steganography means "concealed writing". Stegano means "protected or covered", and graphy means "to write". The purpose of steganography is to hide the presence of communication while the purpose of cryptography is to make the communication incomprehensible by modifying the bit streams using secret keys. The advantage of steganography, over cryptography is that the attackers are not attracted towards communicating messages between sender and receiver while the encrypted messages attract the attackers. Steganalysis is a method of detecting the message hidden in a cover media and to extract it. Changes will be apparent in the statistical property of image if the secret message bits are inserted in image. The strength of the steganography is measured by steganalysis. RS steganalysis is one of the most reliable steganalysis which performs statistical analysis of the pixels to successfully detect the message hidden in the image. However, steganography method to detect the presence of secret message by RS attack/analysis is difficult in case of colour images. Retention of visual quality of the image is also imperative. It is worth to note that genetic algorithm optimizes security and also the quality of the image [1]. It belongs to class of evolutionary algorithms, which imitates the process of natural evolution. This paper introduces a genetic algorithm based steganography method to protect against the RS attack in 24bit colour images.

## 2 LITERATURE SURVEY

The simplest insertion method in steganography is LSB replacement steganography. In the LSB replacement method, the least significant bit of the pixel values are replaced with the bit values of the message. The method of detecting the secret message hidden in the cover media through steganography is known as steganalysis. Steganalysis methods are of two types, one that attacks only color images or grayscale images and the other which attacks on both color and grayscale images. However, irrespective of the aforementioned type of image, some of the steganalysis methods attack only on LSB embedding, while others attack on different methods which also include LSB embedding. Few of the steganalysis methods suspect the message hidden in the image whereas few other steganalysis methods detect the length of the message hidden in the image.

Arezoo Yadollahpour and Hossein Miar Naimi [2] proposed a steganalysis technique using autocorrelation coefficients in colour and grayscale images. They suggest that insertion of secret message weakens the correlation between the neighbour pixels and thereby enable one to detect the message.

Fridrich et al.[3] proposed an effective steganalysis technique popularly known as RS steganalysis, which is reliable even in the detection of non-sequential LSB embedding in digital images.

Andrew D Ker [4] has proposed a general framework for structural steganalysis of LSB replacement for detection and length estimation of the hidden message. He suggests the use of previously known structural detectors and recommended a powerful detection algorithm for the aforementioned purpose.

Tao Zhang and Xijian Ping [6] have proposed a steganalysis method for detection of LSB steganography in natural images based on different histogram. This method ensures reliable detection of steganography and estimate the inserted message rate. However, this method is not effective for low insertion rates.

Fridrich and Goljan [7] considered many steganalysis techniques and proposed a steganalysis technique based on image's biplanes correlation. They state that LSB plane can be estimated from 7 planes out of 8 planes in a pixel of the image. They feel that performance of the suggested steganalysis method reduces as the LSB plane's content is further randomized.

Kong et al. [8] proposed a new Steganalysis approach based on both complexity estimate and statistical filter. It is based on the fact that the bits in the LSB plane are randomised when secret bits are hidden in LSB plane.

Amirtharajan et al. [9] proposed a novel and adaptive method for hiding the secret data in the cover image with high security and increased embedding capacity. They feel that by using this method the receiver do not require the original image to extract the information.

Umamaheswari et al. [10] proposed analysis of different steganographic algorithms for secure data hiding. They recommend compressing the secret message and encrypting it with receiver public key along with the stego key. They have analyzed different embedding algorithms and used cryptographic technique to increase the security.

## 3 GENETIC ALGORITHM BASED STEGANOGRAPHY METHOD

Genetic algorithm based steganogrpahy method incorporates simple LSB embedding technique to hide the data in an image. Each pixel in a 24bit colour image is represented by three bytes where each byte represents the intensity of the three primary colours namely red, green, and blue (RGB), respectively. The data is hidden randomly in the LSB of each byte of the pixels. It is achieved by converting ASCII value of the data into binary format and the bits are hidden in the image by replacing the LSB of the pixel value. The image considered for hiding secret data is a cover image and stego image is obtained by hiding the secret message in a cover image. This research work elucidates the implementation of genetic algorithm to protect the secret data against RS attack in colour images.

RS steganalysis classifies block flipping into three types. They are positive flipping $F_1$, negative flipping $F_{-1}$, and zero flipping $F_0$. RS steganalysis analyses three primary colours namely red, green and blue individually for colour images. Initially, the image is divided into several blocks. Subsequently, flipping functions such as positive flipping and negative flipping are applied on each block of pixels. Later, the variations between original and flipped blocks are calculated. Based on the variation results, the blocks are categorized into regular and singular groups. Let $R_M$ denotes relative number of regular group and $S_M$ denote relative numbers of singular groups. According to the statistical hypothesis of the RS steganalysis method in a typical image, the expected value of $R_M$ is equal to that of $R_{-M}$, and the same is true for $S_M$ and $S_{-M}$:

$$R_M \cong R_{-M} \text{ and } S_M \cong S_{-M}$$

With application of positive flipping, $R_M$ denotes regular group and $S_M$ is singular group. Similarly, $R_{-M}$ and $S_{-M}$ are regular and singular group when negative flipping is applied. The difference between regular groups, $R_M$ and $R_{-M}$ and the difference between singular groups, $S_M$ and $S_{-M}$ increases with the increase in length of the secret message.

In this method, the sender selects and reads an image of size 256x256. The data required to be hidden in the image is converted into bit streams by taking each character of text message and representing their 8 bit binary values from their ASCII code. The data is embedded in the image using LSB embedding technique. The genetic algorithm approach is used to find the best adjustment matrix to protect against RS attack.

Fig 1 depicts the flow of implementation of genetic algorithm based steganography method. Initially, cover image and secret message are read. Secret message is then hidden in the cover image using LSB embedding technique. A stego image is obtained after embedding secret message. The stego image is divided into 8x8 blocks and is labelled by calculating the variations of blocks before flipping and after flipping. During this process, the blocks are categorized into four variables. The variables are based on occurrence of regular group and singular group when positive flipping is used and the occurrence of the regular group and singular group when negative flipping is used. This process is carried out individually for red, green and blue colours. The comparison with the original image shows an increase in certain values of the stego image. The RS attack is therefore able to detect the changes in the values. The genetic algorithm described below is used to decrease the variation in the value of the variables in order to protect against the RS attack.

The genetic algorithm optimizes the image quality and security of the data. Each pixel in a block is considered as a chromosome. Some chromosomes are considered for forming an initial population of the first generation in genetic algorithm. Several generations of chromosomes are created to select the best chromosomes by applying the fitness function to replace the original chromosomes. Reproduction randomly duplicates some chromosomes by flipping the second or third lowest bit in the chromosomes. Several second generation chromosomes are generated. Crossover is applied by randomly selecting two chromosomes and combining them to generate new chromosomes. This is done to eliminate more duplication in the generations. Mutation changes the bit values in which the data bit is not hidden and exchanges any two genes to generate new chromosome. Once the process of selection, reproduction and mutation is complete, the next block is evaluated. The fitness function enables to optimize the value through several iterations. Fitness is calculated by the probability of regular and singular groups when positive flipping and negative flipping is applied. Ultimately, the stego-image undergoes RS analysis and the values between original and stego-image are compared.

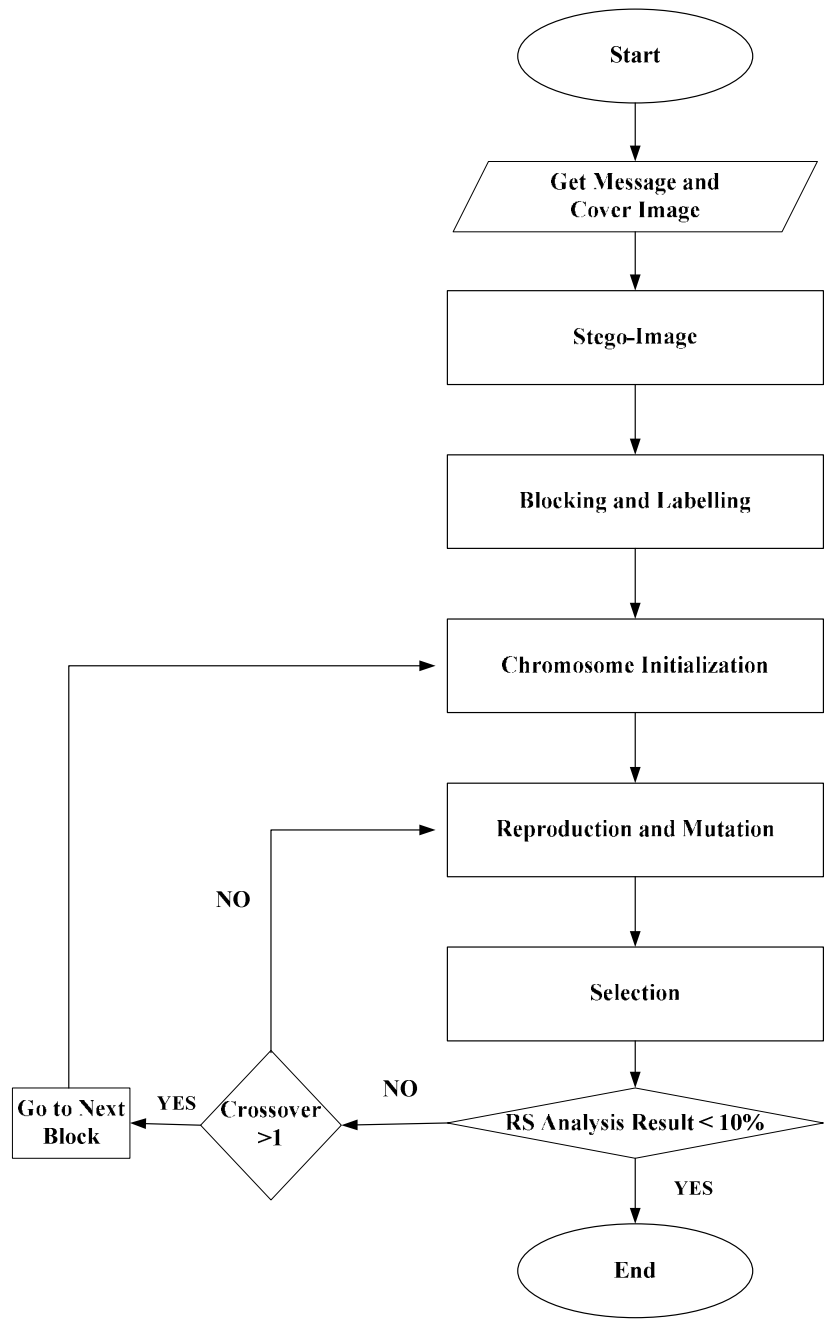

Fig. 1 Procedure of implementation of genetic algorithm based steganography method

## 4 Experimental Results

Fig. 2 depicts two cover images. Fig 2 (a) is a standard Lena image and Fig 2 (b) is a real time image which is captured using a digital camera. It is apparent from the two figures that real time image has more noise when compared to standard image. Security issues are more in real time images due to the presence of noise. Therefore, it is vital to eliminate noise using filters. However, application of apt choice of filters also is a challenge since it is difficult to analyze the type of noise. Further, the quality of image can be measured using methods such as average absolute difference(AAD), mean square error(MSE), laplacian mean square error(LMSE), peak signal to noise ratio(PSNR), normalized cross correlation(NCC) etc.

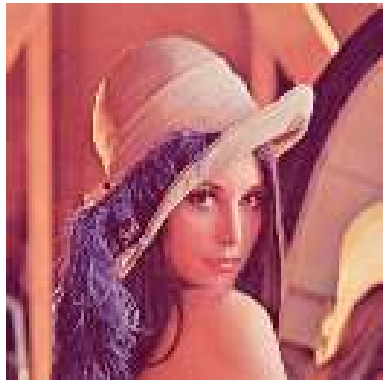
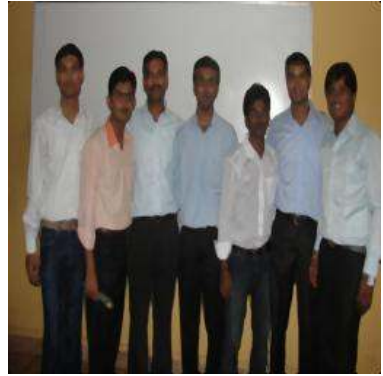

(a) Lena[15]     (b) Friends

Fig. 2 Cover Images

Fig 3 illustrates the results of RS analysis for real time image.

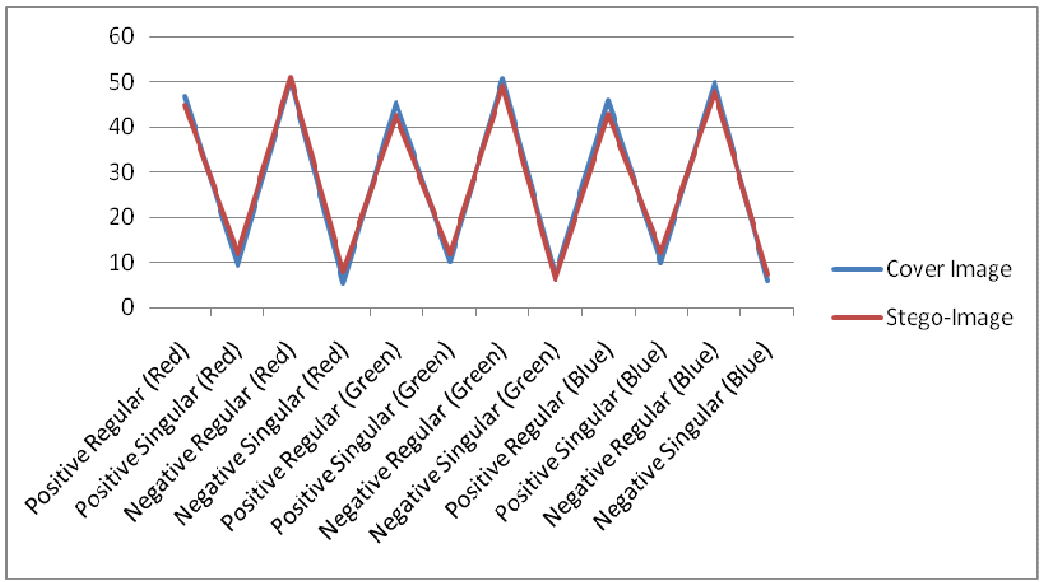

Fig. 3 Percentage of regular and singular groups of stego image and original image

Fig 3 indicates that the difference in percentage of regular and singular groups (both positive and negative) between cover image and stego image is less. Hence, RS steganalysis is not reliable. Therefore, it is difficult to detect the presence of secret message in a stego image.

The comparison between cover image and stego-image of lena.png and friends.jpg are shown in the Table I. The results indicate the presence of noise in real time image. Table further infers that the stego-image is of better quality if AAD, MSE, LMSE value is less while the PSNR value is high and NCC value is approximately equal to 1.

TABLE I

Image Quality Measurement

| Image | AAD | MSE | LMSE | NCC |
|---|---|---|---|---|
| Lena | 0.350 | 0.434 | 0.003 | 0.99 |
| Friends | 0.489 | 0.676 | 0.008 | 0.99 |

Table II depicts the results of RS Analysis for the standard cover image. The table infers that the difference between $R_M$ (Positive Regular) and $R_{-M}$ (Negative Regular) are less than 10%. Further, the difference between $S_M$ (Positive Singular) and $S_{-M}$ (Negative Singular) is also less than 10%. It indicates that the image is secured.

TABLE II

RS Analysis of Original Image (Lena)

|  | Red | Green | Blue |
|---|---|---|---|
| Positive Regular | 34.1888 | 33.3984 | 34.7443 |
| Positive Singular | 19.577 | 19.5984 | 19.9097 |
| Negative Regular | 35.4187 | 34.5093 | 35.3455 |
| Negative Singular | 16.3574 | 15.8447 | 16.7419 |

The Table III shows the RS Analysis for the standard stego image. In stego image also the difference between $R_M$ and $R_{-M}$, $S_M$ and $S_{-M}$ are less than 10%. The comparison of results from Table II and Table III indicates that difference between regular groups and singular groups is less than 10% despite the presence of secret message in stego image. Thus, it is difficult for the RS analysis to detect the secret message hidden in the image.

TABLE III

RS Analysis of Stego Image (Lena)

|  | Red | Green | Blue |
|---|---|---|---|
| Positive Regular | 35.1034 | 34.2764 | 35.6985 |
| Positive Singular | 20.6045 | 21.6747 | 20.4263 |
| Negative Regular | 39.3943 | 37.5215 | 38.3027 |
| Negative Singular | 15.3208 | 16.802 | 15.2487 |

Fig. 5 depicts the images with having hidden secret message. However, from the visual perspective, the quality of the image is maintained and that visual artefact is not introduced

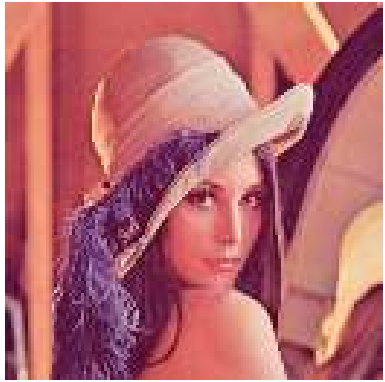
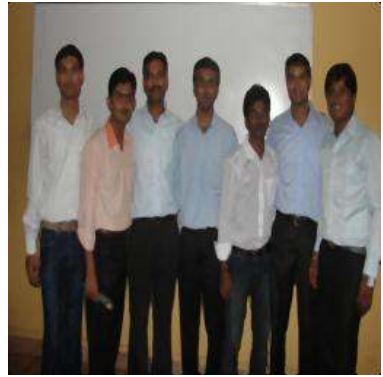

(a) Lena                                         (b) Friends

Fig. 5 Stego Image

## 5   CONCLUSION

Steganography is the art of secret communication. It is the science of hiding a message in such a way that only sender and recipient are aware of existence of the message. The main advantage of steganography is that it will not attract the attackers. The RS analysis is one of the strongest steganalysis, which detects the secret message by the statistical analysis of pixel values.

The objective of this paper is to establish a highly RS-resistant secure model with steganography method using Genetic algorithm. It enables to achieve security and enhance image quality. In this method, the pixel values of the stego image are modified by the genetic algorithm to retain their statistical characteristics. Thus, it is difficult to detect the existence of the secret message by the RS analysis. Further, implementation of this approach enhances the visual quality of the stego image. Nevertheless, as the length of the secret message increases, the probability of detection of secret message by RS analysis also increases. However, our future work focus upon the improvement in embedding capacity and further improvement in the efficiency of this method.